\documentclass{article}
\usepackage{spconf,amsmath,graphicx, booktabs}
\usepackage{amssymb, multirow, color,cite}
\DeclareMathOperator*{\argmin}{argmin}
 \usepackage[colorlinks,
             linkcolor=red,
             anchorcolor=blue,
            citecolor=green]{hyperref}
\usepackage[numbers,sort&compress]{natbib}

\title{UNIVERSAL EFFICIENT VARIABLE-RATE NEURAL IMAGE COMPRESSION}
\twoauthors
 {Shanzhi Yin, Chao Li, Youneng Bao, Yongsheng Liang}
	{Harbin Institute of Technology, Shenzhen
}
 {Fangyang Meng, Wei Liu}
	{Peng Cheng Laboratory
}

\begin{document}

\maketitle

\begin{abstract}
Recently, Learning-based image compression has reached comparable performance with traditional image codecs(such as JPEG, BPG, WebP). However, computational complexity and rate flexibility are  still two major challenges for its practical deployment. To tackle these problems, this paper proposes two universal modules named Energy-based Channel Gating(ECG) and Bit-rate Modulator(BM), which can be directly embedded into existing end-to-end image compression models. ECG uses dynamic pruning to reduce FLOPs for more than 50\% in convolution layers, and a BM pair can modulate the latent representation to control the bit-rate in a channel-wise manner. By implementing these two modules, existing learning-based image codecs can obtain ability to output arbitrary bit-rate with a single model and reduced computation. 
\end{abstract}
\section{Introduction}
\label{sec:intro}
Image compression is a fundamental technology in signal processing and computer vision. It reduces the required bits for image transmission and storage while maintains its reconstruction quality as much as possible. In recent years, many learning-based image compression methods have achieved the state-of-the-art performance comparing to traditional image codecs\cite{Minnen2018,Cheng2020,Chen2021}. However, there are still some challenges  for its practical deployment.

With a predefined trade-off factor, bit-rate and reconstruction quality are fixed for a single trained model. Various requirements needs a potentially large amount of  models and corresponding storage budget.To tackle this disadvantage, quantization steps \cite{Ma2020}, decomposition methods  \cite{8531758,Zhang2019}or trade-off factor \cite{8977394,Choi2019,jia2021rate} are leveraged to obtrain the variable-rate ability. However, these rate-control methods  involve many modifications in original architectures and is hard to be adopted by existing models.
\begin{figure}[t]

    \centering
    \includegraphics[width=8.5cm]{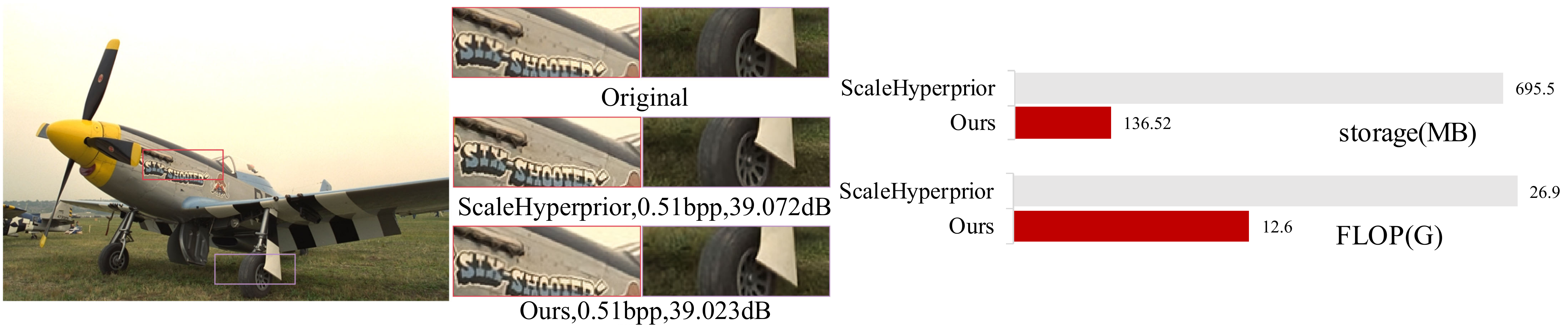}
    \caption{Visualization of reconstructed images from Kodak dataset and corresponding storage\&computation budget}
    \label{fig:0}

\end{figure}

\begin{figure*}[t]

    \centering
    \includegraphics[height=6cm, width=15cm]{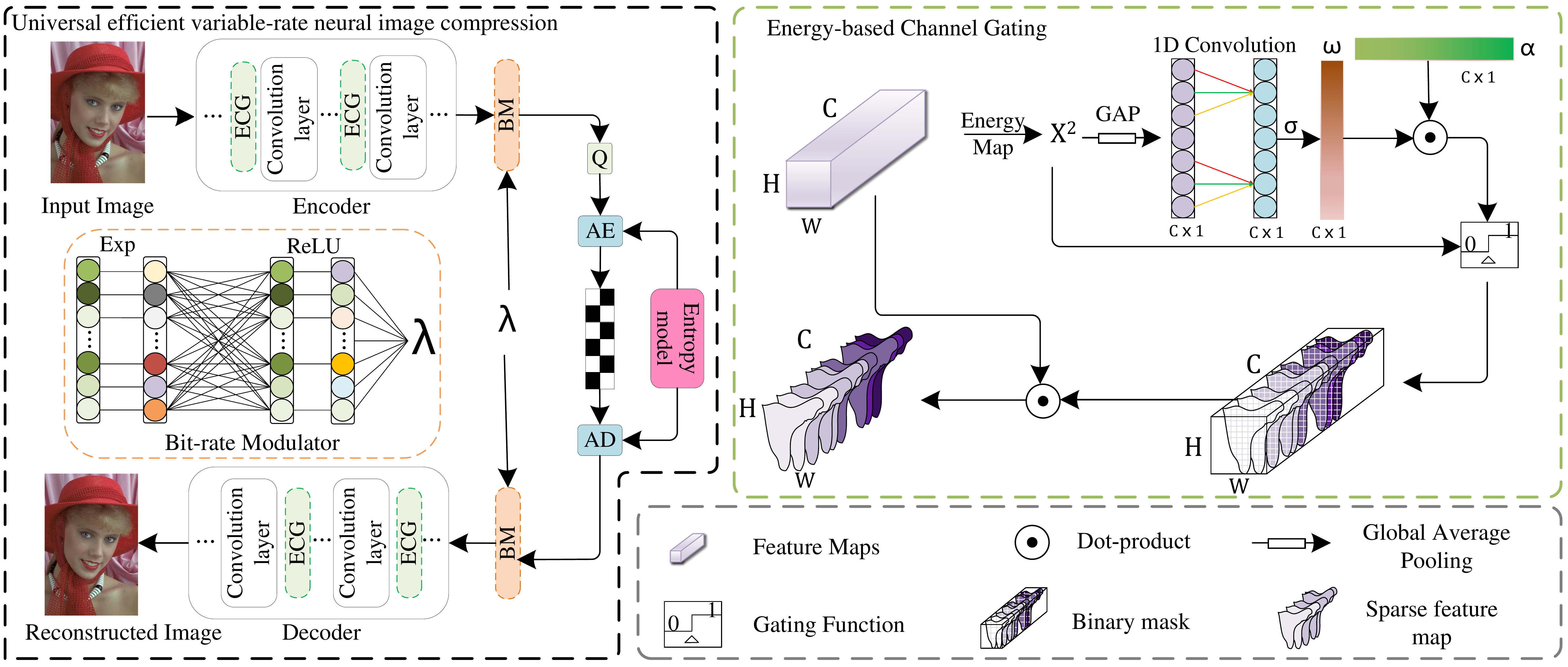}
    \caption{Architecture of universal variable-rate efficient compression model.}
    \label{fig:1}

\end{figure*}

Due to their complex network architectures, computational cost in learning-based compression models is relatively high . In addition, advanced neural network modules are introduced to further boost their performance\cite{Balle2016,Theis2017,Chen2021,Cheng2020,Ma2020} . Heavy computational burden is unfriendly to circumstances like portable electrical devices or edge computing. To tackle this disadvantage,  Johnston et al.\cite{johnston2019computationally}proposed a comprehensive evaluation on compression models, combing  bit-rate, distortion and computation efficiency. Guo et al.\cite{guo2021cbanet}design a complexity adaptive module to replace the decoder. However, the computational complexity levels are predefined in these models, arbitrary complexity is not achieved.

Under real world circumstances, the computation capacity of a given device is usually fixed, but the available transmission bandwidth and reconstruction requirements are varies. To deal with such situation and corresponding challenges of learning-based image compression, this paper proposes a universal variable-rate efficient method for neural image compression(NIC).

Fig.\ref{fig:1} illustrates the architecture of the model. We achieve bit-rate flexibility and reduce computational complexity by embedded  two portable modules i.e. Energy-based channel gating module and Bit-rate modulator into the existing NIC networks. Energy-based channel gating modules are inserted before convolution layers to introduce sparsity, it uses simple 1-D convolution to generate channel-wise threshold and implements dynamic pruning on input feature map. By tuning the learnable adjustment vector, the compression model can achieve arbitrary computational complexity in convolution operations. Bit-rate modulator regards trade-off factor as the input to generate channel-wise multiplier with two full-connected layers only in a plug-in manner. By channel-wisely product the latent representation of pretrained fixed-rate model with the output of the bit-rate modulator, it can modulate the compression bit-rate outside the entropy coding process and generate arbitrary bit-rate. To end-to-end optimize the proposed compression model, we formulate a comprehensive optimization problem including bit-rate, distortion and computational complexity. We implement our method on three classical NIC models and conduct comprehensive the experiments to prove the effectiveness of our method.Visual results and efficiency comparison are shown in Fig.\ref{fig:0}. There is no obvious qualitative degredation on the reconstructed image, while the storage and computation saving is quite impressive.

In this paper, our main contribution is to present a method to built variable-rate and low-computation image compression based on existing models. We design a dynamic feature map pruning module that can reduce the FLOP of NIC models up to  $3\times$ and we also design a the simple but effective plug-in bit-rate modulator that can obtain arbitrary  bit-rate of NIC models. We use modules mentioned-above to form universal efficient variable-rate NIC and generalize it to three existing models. By solving a comprehensive optimization problem, experiments show  that our method can achieve continuous bit-rate flexibility in a single model with around half of the original computation.

\section{Proposed Method}
\label{sec:Method}
\subsection{Energy-based channel gating}
\label{sec:cg}
Sparsity is an effective way to reduce computational complexity. In convolution operations, inputs of different intensities have different influence on the results, sparsity can be introduced by pruning the inputs with less significance, which will be judged by energy-based channel gating module(ECG). Inspired by \cite{NEURIPS2019_68b1fbe7}, ECG uses learnable dynamic feature map pruning with channel-wise threshold. The energy of the input is evaluated within each channel and between its neighbour channels\cite{9156697} to obtain threshold for every channel.

As shown in Fig.\ref{fig:1}, given the input $x$, the energy map $x^2$ is first adaptive average pooled(GAP) to obtain the intensity information of a single channel, then 1-D convolution with adaptive kernel size is conducted following\cite{9156697} to integrate the intensity information of neighbour channels, the convolution result is activated by Sigmoid function to form a importance vector $\omega$:
\begin{equation}\label{importancevector}
\omega = \sigma\{C1[P(x^2)]\} 
\end{equation}
in which $P$ is adaptive average pooling, $C1$ is 1-D convolution, $\sigma$ is Sigmoid activation. The importance vector then dot-products a learnable adjustment vector $\alpha$ to form a learnable threshold $th=\omega\odot\alpha$.  A binary mask is generated by gating the energy map $x^2$ through this channel-wise threshold $th$, the gating function can be described as:
\begin{equation}\label{gt}
gt(x) =
\begin{cases}
1 & \text{if } x \ge 0,\\
0 & \text{if } otherwise,\\
\end{cases} 
\end{equation}
For this gating function is not differentiable, Sigmoid function $\sigma(x)=\frac{1}{1+e^{-\epsilon x}}$ is used to replace it during the training process\cite{NEURIPS2019_68b1fbe7}. The final  output of ECG $x_{mask}$  is the original input $x$ dot-product the binary mask:
\begin{equation}\label{xmaks}
x_{mask} = x\odot gt(x^2-\omega\odot \alpha)
\end{equation}$x_{mask}$  is the sparse version of the original input $x$ and the computation can be reduced with its sparsity.

\subsection{Bit-rate modulator}
\label{sec:bm}
Following the work in \cite{cui2021asymmetric}, we try to affect the R-D performance of the compression model through the quantization process. Instead of learning a gain matrix\cite{cui2021asymmetric}, we use bit-rate modulator shown in Fig.\ref{fig:1} to implement the bottleneck scaling  in a channel-wise and plug-in manner. 

To obtain rate-flexibility in a convenience but effective way, we design the bit-rate modulator(BM) to be a simple and portable module with two full-connected layer only\cite{8977394}. It maps a trade-off factor $\lambda$ into a vector which has the same channel number as the entropy coding process:
\begin{align}\label{bm}
f_1(\lambda) = ReLU(\mathbf{H^{(1)}}\lambda+\mathbf{b^{(1)}})\\
bm(\lambda) = exp(\mathbf{H^{(2)}}f_1+\mathbf{b^{(2)}})
\end{align}
in which the exponential mapping before the final output is to maintain the positive value of the vector and expand the dynamic range of the modulator.

During the compression process, the latent representation $y$ channel-wisely products the $bm(\lambda)$ after the encoder to obtain the modulated latent $y_{mod}$. While before the decoder, $y_{mod}$ products the inverse module $ibm(\lambda)$ to restore the original feature map.
\subsection{Implementation on existing compression models}
\label{sec:net}
To illustrate the effectiveness of our methods, we implement it on several classical neural image compression models\cite{balle2018variational,Minnen2018}
denoted as  ScaleHyperprior model,  MeanscaleHyperprior model and JointAutoregressive model respectively. 

\begin{figure}[t]

    \centering
    \includegraphics[width=8.5cm]{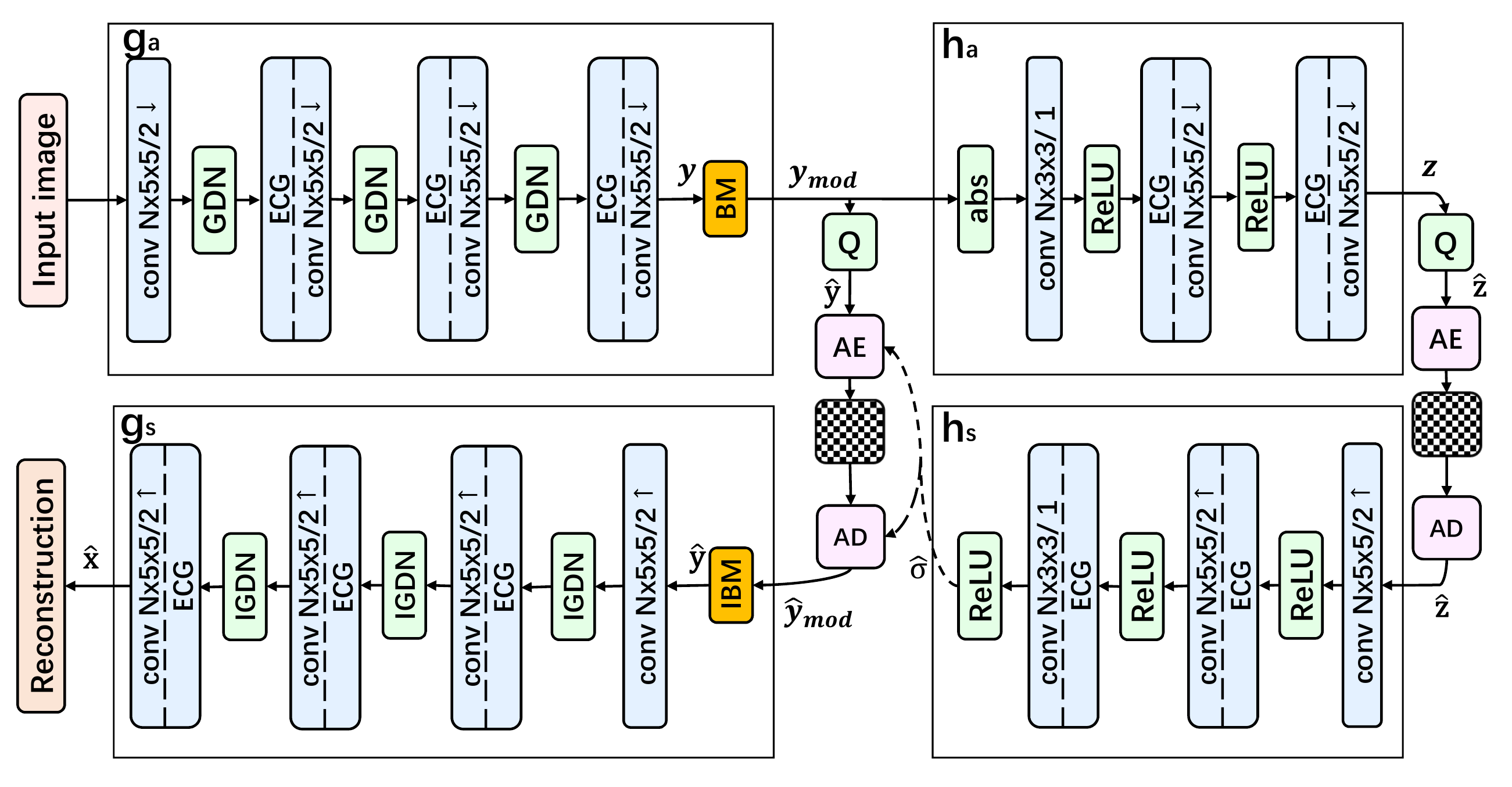}
    \caption{Implementation on ScaleHyperprior model}
    \label{fig:3}

\end{figure}
In this section, we use ScaleHyperprior model as an example to show the implementation details and optimization strategies of our method. 

As shown in Fig.\ref{fig:3}, we embedded totally 10 ECG modules with convolution layers  to reduce the computational complexity, while 4 convolution layers remain unchanged. These 4 layers receive the input image or feature map of main codecs and hyper codecs respectively. Keeping these layers can maintain the receptive ability of the codecs to avoid dramatic performance degradation. 
Bit-modulator is inserted as the last layer of the main encoder and the first layer of the main decoder to modulate the feature map for  ScaleHyperprior entropy model. The feature map of hyper codecs remains unchanged, for the bit-rate produced by it is relatively low\cite{balle2018variational}and the predicted scales are significant for keeping the ScaleHyperprior model as accurate as possible.

After adopting our method into the original ScaleHyperprior model, it needs to be trained in end-to-end manner. Similar to most neural image compression model, two loss terms rate $R$ and distortion $D$ need to be optimized.
As the actual arithmetic coding is bypassed\cite{Balle17a}, the rate term is given by the cross entropy of the estimated distribution of $y$ and the its actual distribution:
\begin{equation}\label{rate}
R(\hat{y};\theta,\phi, \xi,\lambda) = \mathbb{E}_{\hat{y}\sim p_{y}}\{log_2q_{y}[Q(y\odot bm(\lambda))]\}
\end{equation}in which $\theta$ , $\phi$  ,$\xi$ denote the parameter of codecs, ECG Modules, BM modules respectively and $\lambda$ is the trade-off factors as the input of the BM module, $Q$ represents the quantizaiton process, $p$ and $q$ are actual and predicted distribution of image data respectively.
In our work, the distortion metric is the mean square error measured on the test set:
\begin{equation}\label{distortion}
D(x,\hat{x};\theta,\phi, \xi,\lambda) = \mathbb{E}_{x\sim p_{x}}[||x-\hat{x}||^2]
\end{equation}
The R and D loss term are weighted-summed by a trade-off factor $\lambda$. When we try to obtain the rate flexibility through multiple R-D trade-offs, the variable-rate compression optimization problem can be form as:
\begin{equation}\label{lossvr}
\mathop{\argmin_{\theta,\phi, \xi,\lambda}\sum_{\lambda \in \Lambda}[R(\hat{y};\theta,\phi, \xi,\lambda)+\lambda D(x,\hat{x};\theta,\phi, \xi,\lambda)]}
\end{equation}in which $\Lambda$ is the set of all possible values of $\lambda$, i.e $\Lambda=\{\lambda_0,\lambda_1\cdots \lambda_n\}$

Unlike traditional neural image compression, our method introduce sparsity through learnable ECGs which need to be optimized during the training process as well. In ECG, the learnable adjustment vector $\alpha$ affect the final gating threshold $th$, larger the $\alpha$ is, higher the final threshold on each channels will be, and the output feature map of ECG will be sparser. During the training process, we set a relatively high target value $\alpha_t$ for the learnable adjustment vector and take the mean square error of present $\alpha$ and $\alpha_t$ as an additional loss term. With this loss term being optimized, we can ensure that, the sparsity of the ECG output can gradually reach an expected level and overall optimization under such multiple trade-offs can be achieved. The final optimization problem can be formulated as:
\begin{equation}\label{loss}
\mathop{\argmin_{\theta,\phi, \xi,\lambda}\sum_{\lambda \in \Lambda}[R+\lambda D+\gamma \sum_{i=1}^n(\alpha_n-\alpha_t)^2]}
\end{equation}in which $\alpha_i$ represents the $i$th ECG module, $\gamma$ is the trade-off factor for computation effeciency and $R$, $D$ are defined as equation(\ref{rate})and(\ref{distortion}). When tuning the $\gamma$, arbitrary computational comlexity can be achieved, so that the model can be adopted on various computational environments.

\section{Experiments}
\label{sec:Experiments}
The NIC\_dataset \footnote{The NIC\_Dataset can be accessed at https://www.bitahub.com/dataset, which is opened as a public dataset.} is used for training, which contains 607,714 256x256 patches cropped from the 1,600 original images and the 2x, 4x down-sampled versions using bicubic interpolation. We adapted CompressAI\cite{begaint2020compressai} re-implementation of  neural compression models and followed their training parameters settings and training strategies. All trained models are evaluated on Kodak dataset.
\subsection{Ablation study}

\begin{table*}[t]
  \centering
  \caption{The PSNR and FLOP reduction of models with ECG module}
    \begin{tabular}{c|c|c|c|c|c|c|c|c|c}
    \toprule
    \multirow{2}{*}{Model} & \multirow{2}{*}{Performance} & \multicolumn{8}{c}{Quality} \\
\cline{3-10}          &       & 1     & 2     & 3     & 4     & 5     & 6     & 7     & 8 \\
    \hline
    \multirow{2}{*}{ScaleHyperprior} & PSNR drop(\%) &  0     &  0.346     & 0.228      &   0.269    &     0.336  & 0      & 0      & 0.216 \\
    \cline{2-10}
          & FLOP reduction &   \textbf{2.54$\times$}    & \textbf{2.86$\times$}      &\textbf{ 2.60$\times$}      & \textbf{2.54$\times$}      &\textbf{2.54$\times$}       &\textbf{2.07$\times$}       &\textbf{2.14$\times$}       &\textbf{2.03$\times$} \\
    \hline
    \multirow{2}{*}{MeanscaleHyperprior} & PSNR drop(\%) & 0.39     &  0.22     &   0.77    & 0.69     &   0.37    & 0.61     &    0.71  & 0.78 \\
    \cline{2-10}
          & FLOP reduction &   \textbf{2.34$\times$}    & \textbf{2.50$\times$}      &\textbf{2.56 $\times$}      & \textbf{2.68$\times$}      &\textbf{2.33$\times$}       &\textbf{2.12$\times$}       &\textbf{2.12$\times$}       &\textbf{2.24$\times$} \\
    \hline
    \multirow{2}{*}{JointAutoregressive} & PSNR drop(\%) &  0.207     &0.335       & 0.437      & 0.807     & 0.465      &0.150       &0.354       &0.553  \\
    \cline{2-10}
          & FLOP reduction &\textbf{2.43$\times$}  &\textbf{2.67$\times$}  &\textbf{2.48$\times$}  &\textbf{2.23$\times$}       & \textbf{2.29$\times$}      &\textbf{2.02$\times$}       &\textbf{2.06$\times$}       &\textbf{2.02$\times$}  \\
    \bottomrule
    \end{tabular}%
  \label{tab:addlabel}%
\end{table*}%

\begin{figure}[t]
      \centering
      \centerline{\includegraphics[width=8.5cm]{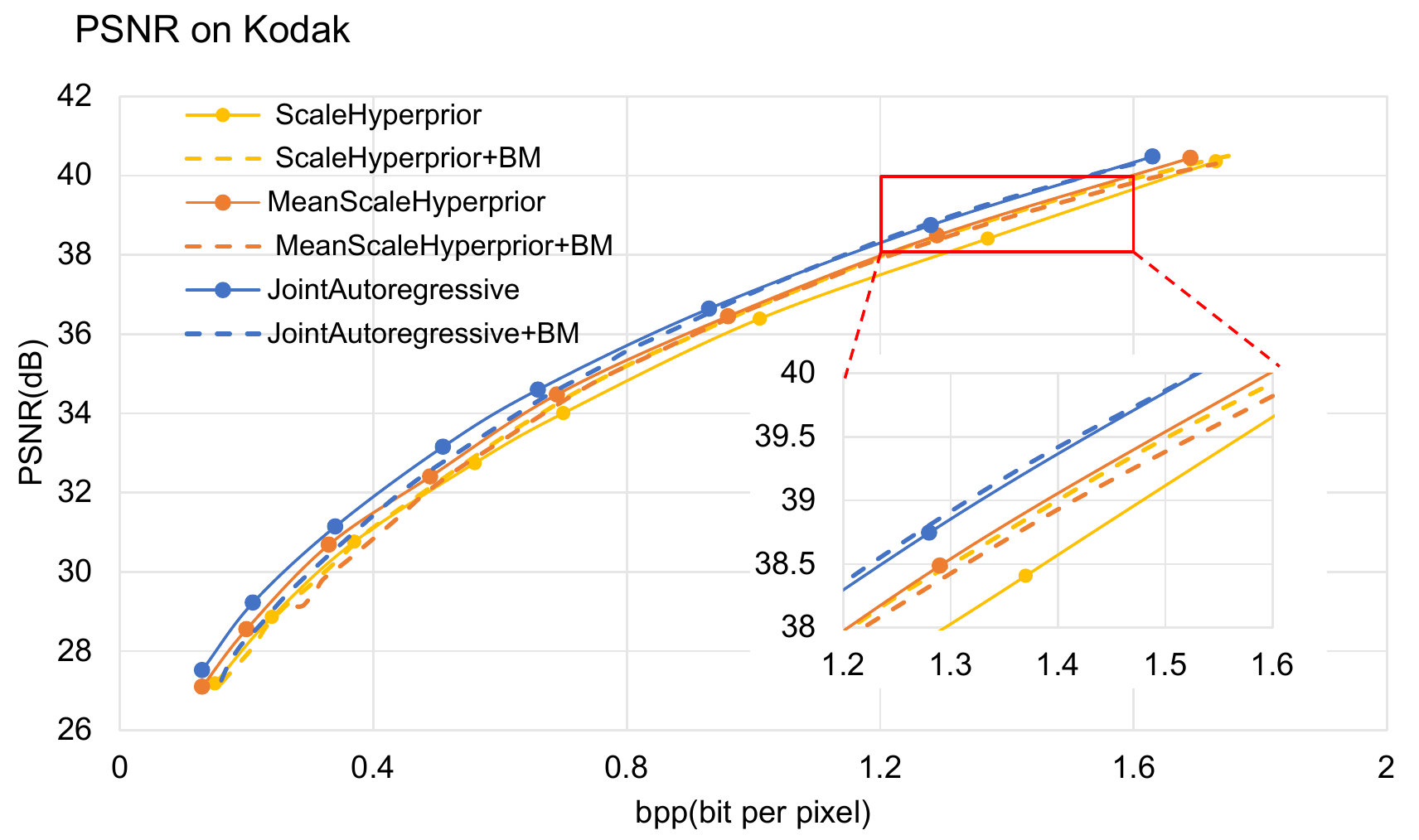}}
    \caption{Bit-rate flexibility on three classical neural image compression model}
    \label{fig:4}

\end{figure}
We first evaluate the effectiveness of our ECG module by embedding it into  ScaleHyperprior, MeanscaleHyperprior and JointAutoregressive models.$\gamma$ and $\alpha_t$ was empirically set to be 0.0001. We follow the 8 different trade-off factors in\cite{begaint2020compressai} for training the original models. When embedded with ECGs, we observe that the bit-rate of  efficient models fluctuate around their original bit-rate. For fair comparison, we fine-tunned the trade-off factor for efficient models so that their bit-rates are the same (accurate to two decimal places) as their original counterparts. PSNR and FLOP reduction are shown in Table.\ref{tab:addlabel}. We can see that the FLOP reduction of more than $ {2\times}$ can be achieved in three neural compression models with very slight PSNR degradation around 0.5\% and no more than 1\%.

Then we evaluate the effectiveness of our BM module. We use pre-trained highest quality original models and fine-tuned them with BM module. The R-D performance is shown in Fig.\ref{fig:4}. We can see that continuous rate flexibility can be achieved in three architecture with only one trained model. Compared with previous 8 models of 695.5MB,1125.6MB,
1916MB, one single model only takes 135MB,201MB and 311MB storage, saving up to 91.7\%.
\subsection{R-D and efficiency performance}
\begin{figure}[t]
      \centering
      \centerline{\includegraphics[width=8.5cm]{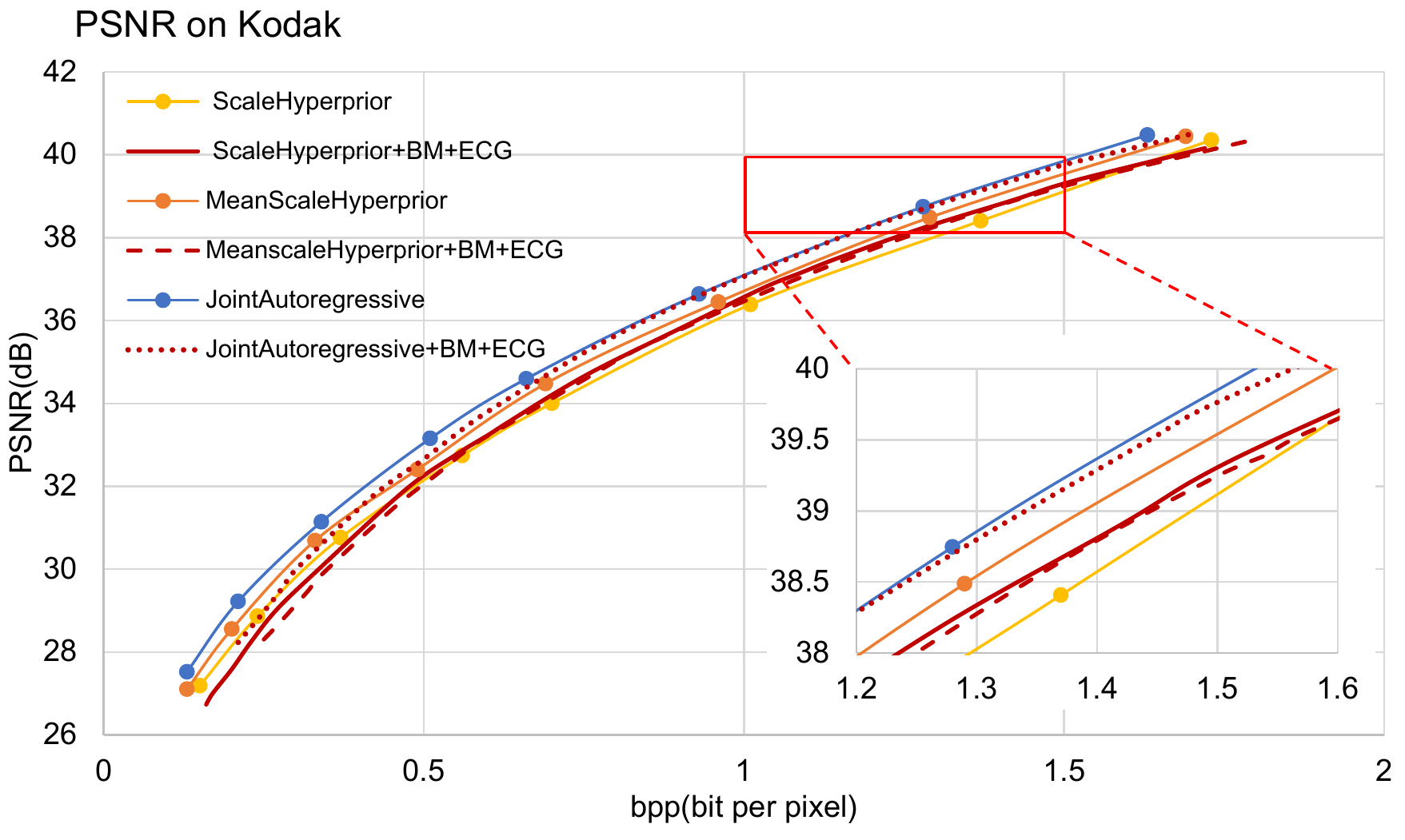}}
    \caption{Performance of universal variable-rate efficient models}
    \label{fig:5}

\end{figure}
We implement both ECG and  BM modules to form variable-rate and low- computational complexity efficient compression model on above-mentioned three architectures. We use ECG embedded models with highest quality as pre-trained models and fine-tuned them with BM module. The R-D  performance is shown in Fig.\ref{fig:5}. We can see that three variable-rate efficient models are able to approach the performances of the original models. The models can achieve  sparsity around \textbf{0.5} in convolution operations and achieve storage saving of \textbf{80.5\%},\textbf{84.5\%} and \textbf{91.7\%} respectively.

\section{Conclusion}
\label{sec:Conclusion}
We proposed  two simple modules, energy-based channel gating module and bit-rate modulator, which can be adopted into most of the existing models. Continuous bit-rate flexibility can be achieved with reduced FLOP. In our experiment, existing compression models can generate arbitrary bit-rate and reconstruction quality in a single trained model with 80-90\% storage saving and approximately 2-3$\times$FLOP saving, which makes it easier for their pratical deployment. Protential future work includes better joint training strategies for overall performance and  adaptive computational cost in single model.
\bibliographystyle{IEEEbib}
\bibliography{refs}

\end{document}